\documentclass[twocolumn]{aastex631}

\usepackage{lmodern}
\usepackage{latexsym}
\usepackage{amsmath}
\usepackage{amssymb}
\usepackage{amsbsy}
\usepackage{amsthm}
\usepackage{amsfonts}
\usepackage{mathrsfs}
\usepackage{bm}
\usepackage{sansmath}
\usepackage{relsize}
\usepackage{caption2}
\usepackage{graphicx}
\usepackage[utf8]{inputenc} % usually not needed (loaded by default)
\usepackage[T1]{fontenc}
\usepackage{epstopdf}

\shorttitle{Self-similar solution of hot accretion flow}
\shortauthors{Mosallanezhad et al.}
\graphicspath{{./}{figures/}}
\begin{document}

\title{Self-similar Solution of Hot Accretion Flow with Thermal Conduction and Anisotropic Pressure}

\correspondingauthor{Liquan Mei}
\email{lqmei@mail.xjtu.edu.cn}

\author[0000-0002-4601-7073]{Amin Mosallanezhad}
\affiliation{School of Mathematics and Statistics, Xi'an Jiaotong University, Xi'an, Shaanxi 710049, PR China}

\author[0000-0003-3345-727X]{Fatemeh Zahra Zeraatgari}
\affiliation{School of Mathematics and Statistics, Xi'an Jiaotong University, Xi'an, Shaanxi 710049, PR China}

\author[0000-0003-3468-8803]{Liquan Mei}
\affiliation{School of Mathematics and Statistics, Xi'an Jiaotong University, Xi'an, Shaanxi 710049, PR China}

\author[0000-0002-0427-520X]{De-Fu Bu}
\affiliation{Shanghai Astronomical Observatory, Chinese Academy of Sciences, Shanghai 200030, China}



\begin{abstract}
We explore the effects of anisotropic thermal conduction, 
anisotropic pressure, and magnetic field strength on the
hot accretion flows around black holes by solving the axisymmetric, 
steady-state magnetohydrodynamic equations. The anisotropic pressure
is known as a mechanism for transporting angular momentum in 
weakly collisional plasmas in hot accretion flows with extremely 
low mass accretion rates. However, anisotropic pressure does not extensively
impact the transport of the angular momentum, it leads to shrinkage of the wind region.
Our results show that the strength of the magnetic field
can help the Poynting energy flux overcomes the kinetic energy flux.
This result may be applicable to understand the hot accretion flow in the 
Galactic Center Sgr A* and M87 galaxy.
\end{abstract}



\section{Introduction} \label{sec:intro}

According to the temperature of the accretion flow, black hole accretion disks can 
be divided into two broad types; cold and hot. Intensive analytical studies and simulations
have been done on hot accretion flows, since the pioneer study 
of \citealt{Narayan and Yi 1994}. It is well known that variety of accreting systems in the universe 
with low mass accretion rate such low-luminosity active galactic nuclei (LLAGNs) as 
well as black hole X-ray binaries in the quiescent and hard states can be explained 
by hot accretion flow model (\citealt{Yuan and Narayan 2014}). Indeed, hot accretion flow 
is an efficient model to describe the accretion flow around majority of nearby galaxies 
with extremely low-luminosity including our galactic center, Sagittarius $\mathrm{A^*}$ 
(Sgr $\mathrm{A^*}$) and M87. 

Based on observations of black hole X-ray binaries and LLAGNs, outflow (wind and jet) 
can be launched from hot accretion flow (\citealt{Tombesi et al. 2010, Tombesi et al. 2014}; 
\citealt{Crenshaw and Kraemer 2012}; \citealt{Wang et al. 2013}; \citealt{Cheung et al. 2016}; 
\citealt{Ma et al. 2019}). Wind can push the gas around the black hole outward and 
affectively change the black hole accretion rate in hot accretion systems. Wind is not only 
a key ingredient of accretion flows which can impact on their dynamics and structure, but  
feedback of the wind on the surrounding medium can also play a major role in the formation and 
evolution of galaxies (\citealt{Fabian 2012}; \citealt{Kormendy and Ho 2013}; 
\citealt{Naab and Ostriker 2017}). 

Three different wind-launching mechanics have been proposed, including 
thermally driven (\citealt{Begelman et al. 1983}; \citealt{Font et al. 2004}; 
\citealt{Luketic et al. 2010}; \citealt{Waters and Proga 2012}), magnetically driven 
(\citealt{Blandford and Payne 1982}; \citealt{Lynden-Bell 1996, Lynden-Bell 2003}), 
and radiation-driven wind (\citealt{Murray et al. 1995}; \citealt{Proga et al. 2000}; 
\citealt{Proga and Kallman 2004}; \citealt{Nomura and Ohsuga 2017}). In terms 
of hot accretion flow, the first two mechanisms play a major role in producing wind 
since the radiation is negligible (\citealt{Yuan and Narayan 2014}). To investigate the existence of wind in a hot 
accretion flow, a large number of hydrodynamical (HD) and magnetohydrodynamical 
(MHD) numerical simulations have been carried out so far (e.g., \citealt{Stone et al. 1999}; 
\citealt{Igumenshchev and Abramowicz 1999}; \citealt{Stone and Pringle 2001}; 
\citealt{Hawley et al. 2001}; \citealt{Machida et al. 2001}; \citealt{Pen et al. 2003}; 
\citealt{De Villiers et al. 2003}, 2005; \citealt{Yuan and Bu 2010}; \citealt{Pang et al. 2011}; 
\citealt{McKinney et al. 2012}; \citealt{Narayan et al. 2012}; \citealt{Li et al. 2013}; 
\citealt{Yuan et al. 2012a, Yuan et al. 2012b, Yuan et al. 2015};  \citealt{Bu and Zan 2018};  
\citealt{Inayoshi et al. 2018, Inayoshi et al. 2019}).

Hot accretion flows with very low mass accretion rate are in fact virialy hot and low-density plasmas 
 (\citealt{Narayan et al. 1998}; \citealt{Quataert 2003}; \citealt{Yuan and Narayan 2014}). 
Since the density is very low, the plasma is collisionless, i.e., the Coulomb mean free path 
is many orders of magnitude larger than the system size. Therefore, the electrons should 
move large distance to exchange the energy with other particles. In such a case, 
thermal conduction is important and can significantly 
change the dynamics of the accretion flow. On the other hand, it is well known that the 
magnetorotational instability (MRI) is a mechanism to induce magnetohydrodynamic turbulence 
for driving the angular momentum outward in accretion flows (\citealt{Balbus and Hawley 1991}). 
Consequently, the MHD equations of hot accretion flow should include thermal conduction. 
 Moreover, the ions mean free path is also quite large compare to the larmor radius in hot accretion 
flow with low mass accretion rate. The ions move along the field lines. Therefore, the pressure 
should be anisotropic. We can treat the anisotropic pressure as a perturbation to the ideal MHD and
include it to the MHD equations.

There are several analytical investigations based on the self-similar assumption that  
probe the magnetic field effects in hot accretion flows (\citealt{Akizuki and Fukue 2006}; 
\citealt{Abbassi et al. 2008}; \citealt{Zhang and Dai 2008}; \citealt{Bu et al. 2009}; 
\citealt{Mosallanezhad et al. 2014}; \citealt{Samadi et al. 2017}; \citealt{Bu and Mosallanezhad 2018}; 
\citealt{Deng and Bu 2019}). In practice, the analytical studies are powerful tools to 
understand the physical properties of inflow and wind from accretion flow. 
The reasons are as follows: (1) global numerical simulations of accretion flow 
are expensive and time-consuming because of the complex physics involved 
such as dissipation and magnetic field, (2) analytical techniques are straightforward 
to check the dependency of the results on the variety of physical input parameters.

In the present study, we also adopt radially self-similar assumption and solve 
the MHD equations of hot accretion flow in the vertical direction. We mainly focus on two 
extreme cases of magnetic field, namely weak and strong field. The magnetic field 
strength will be initialized by the usual plasma beta definition as $ \beta_{_0} = 
p_\mathrm{gas} / p_\mathrm{mag} $, where $ p_\mathrm{gas} $ and $ p_\mathrm{mag} $ 
represent the gas and magnetic pressure at the equatorial plane, respectively. 
Our main goal is to study the influence of the anisotropic pressure, magnetic 
field strength and thermal conduction on the properties of the inflow and wind. 
More precisely, we examine how the physical input parameters such as thermal 
conductivity coefficient and initial plasma beta change the inflow and wind regions 
of the hot accretion flow. We also investigate the mass-flux weight properties of 
wind from hot accretion flow in both weak and strong field cases.

The remainder of the manuscript is organized as follows. In Section 
\ref{sec:numerical_method}, the basic equations, physical assumptions, 
self-similar solutions, and the boundary conditions will be introduced. 
The numerical results of weak and strong magnetic field cases will be 
presented in details in Section \ref{sec:results}. Finally, we will provide 
the summary and discussion in Section \ref{sec:summary_discussion}.

\section{Numerical method and Assumptions} \label{sec:numerical_method}

\subsection{Basic Equations} \label{subsec:basic_equations}

The basic equations of the hot accretion flow with extremely low-luminosity 
in the presence of viscosity, magnetic field, anisotropic pressure and thermal 
conduction can be described as, 

\begin{equation} \label{eq:continuity}
  \frac{ \mathrm{d} \rho}{\mathrm{d} t} +\rho \bm{\nabla} \cdot \bm{v} = 0,
\end{equation}

\begin{equation}\label{eq:momentum}
\rho  \frac{\mathrm{d} \bm{v}}{\mathrm{d} t} = - \rho \bm{\nabla} \psi  - \bm{\nabla} p_\mathrm{gas} + \bm{\nabla} \cdot \bm{\mathrm{\sigma}} + \frac{1}{c} \left( \bm{J} \times \bm{B} \right) + \bm{\nabla} \cdot \bm{\Pi},
\end{equation}

\begin{equation}\label{eq:energy}
	\rho \frac{\mathrm{d} e}{\mathrm{d} t} - \frac{p_\mathrm{gas}}{\rho} \frac{\mathrm{d} \rho}{\mathrm{d} t}  \equiv 
	Q^{+} + \bm{\nabla} \cdot \bm{Q}_{c} - \bm{\Pi} : \bm{\nabla} \bm{v},
\end{equation}

\begin{equation} \label{eq:induction}
	\frac{\partial \bm{B}}{\partial t} = \bm{\nabla} \times \left( \bm{v} \times \bm{B} - \frac{4 \pi}{c} \eta \bm{J} \right), 
\end{equation}

\begin{equation} \label{eq:delB}
	\bm{\nabla} \cdot \bm{B} = 0.
\end{equation}

In the above equations,  $ \rho $ is the mass density, $ \bm{v} $ is the velocity, 
$ \psi = - GM / r $ is the Newtonian potential (where $ G $ is the gravitational constant, 
$ M $ is the mass of the central black hole, and $ r $ is the distance from the black hole), 
$ p_\mathrm{gas} $ is the gas pressure, $ \bm{\sigma} $ is the viscous stress 
tensor, $ \bm{J} = c \left( \nabla \times \bm{B} \right) / 4\pi $ is the current 
density (where $ c $ is the speed of light), $ \bm{B} $ is the magnetic field,
$ \bm{\Pi} $ is anisotropic pressure, $ e $ is the gas internal energy, $ Q^+ $ 
is the total heating due to viscous and magnetic dissipation, $ \bm{Q}_{c} $ is 
the thermal conduction, and $ \eta $ is the magnetic diffusivity.
The $ \mathrm{d}/\mathrm{d} t \equiv \partial / \partial t +  \bm{v} \cdot \nabla $ 
denotes the Lagrangian or comoving derivative. The equation of state of the 
ideal gas is assumed as $ p_\mathrm{gas} =  \left( \gamma - 1 \right) \rho e $, 
with $ \gamma = 5/3 $ being the adiabatic index. In fact, in accreting systems 
Maxwell stress associated with MHD turbulence driven by MRI (\citealt{Balbus and Hawley 1998})
can transfer the angular momentum outward. Following the MHD 
numerical simulations, we decompose the magnetic field into a 
large-scale component and a turbulent component. The magnetic field, $ \bm{B} $,
in the above equations represents the large-scale component. The effects of 
turbulence are described by the viscosity terms of the momentum 
and the energy equations. Both components can transport the angular momentum 
outward and produce heat. Following numerical simulations of hot accretion 
flow, \citealt{Stone et al. 1999}; \citealt{Yuan et al. 2012b}, we assume 
the azimuthal component of the viscous stress tensor, $ \sigma_{r \phi} $, 
is the dominant component. This component is written as

\begin{equation}\label{sigma_rp}
        \sigma_{r \phi} =  \rho \nu_{1} r \frac{\partial}{\partial r} \left( \frac{v_{\phi}}{r} \right),
\end{equation}
where $ \nu_{1} $ is the kinematic viscosity coefficient which is evaluated as,

\begin{equation}\label{viscous_stress_tensor}
\nu_{1} = \frac{\alpha_{1}}{ \rho\, \Omega_{\!_{K}}} \left( p_\mathrm{gas} + p_\mathrm{mag} \right),
\end{equation}
where $ \Omega_{\!_{K}} (\equiv GM/r^3)^{1/2} $ is the Keplerian velocity, 
$ \alpha_{1} $ is the viscosity parameter, and $ p_\mathrm{mag} = | \bm{B} |^{2} / (8 \pi) $ 
is the magnetic pressure. The heating rate $ Q^+ $ in equation (\ref{eq:energy}) is 
decomposed into two components, i.e., the viscous heating and the magnetic field 
dissipation heating as,

\begin{equation} \label{heating_rate}
	Q^{+} = Q_\mathrm{vis} + Q_\mathrm{res},
\end{equation}
with
\begin{equation}
Q_\mathrm{vis} = \nabla \bm{v} : \bm{\mathrm{\sigma}},
\end{equation}

\begin{equation}
Q_\mathrm{res} = \frac{4\pi }{c^2} \eta \bm{J}^2.
\end{equation}

Here, the magnetic diffusivity can be set as $ \eta = \eta_{0} / ( \rho \Omega_{\!_{K}} ) 
\left[ p_\mathrm{gas} + p_\mathrm{mag} \right] $ to satisfy the self-similar solutions 
in the radial direction. Following the numerical simulations (e.g., \citealt{Braginskii 1965}; 
\citealt{Balbus 2004}; \citealt{Chandra et al. 2015}), we model the anisotropic pressure 
$ \bm{\Pi} $ with an anisotropic viscosity,

\begin{equation} \label{anisotropic_viscosity}
\Pi = -3\rho \nu_2 \left[ \hat{\bm{b}} \hat{\bm{b}} : \nabla \bm{v} - \frac{\nabla \cdot \bm{v}}{3}\right] 
\left[ \hat{\bm{b}} \hat{\bm{b}} - \frac{\bm{\mathrm{I}}}{3} \right],
\end{equation}
where $ \hat{\bm{b}} = \bm{\mathrm{B}} / | \bm{\mathrm{B}}  | $ is a unit vector in 
the direction of magnetic field, and $ \bm{\mathrm{I}} $ is the unit tensor. The anisotropic 
viscosity coefficient $ \nu_{2} $ is written as, 

\begin{equation} \label{nu_2}
	\nu_{2} = \frac{\alpha_{2}}{ \rho \Omega_{\!_{K}}} \left( p_\mathrm{gas} + p_\mathrm{mag} \right).
\end{equation}

The hot accretion flow is taken to be axisymmetric and steady state ($ \partial/\partial \phi = 
\partial/\partial t = 0 $). We adopt spherical polar coordinates $ (r, \theta, \phi) $ to solve the 
Equations (\ref{eq:continuity})-(\ref{eq:delB}). To avoid complexity due to considering all 
components of magnetic field, in our current analytical study, we assume only toridal component 
of the field as

\begin{equation} \label{magnetic_field_decomposed}
	\bm{B} = B_{\phi} (r,\theta) \bm{e}_{\phi},
\end{equation}
which automatically satisfies $ \nabla \cdot \bm{B} = 0 $ \footnote{In future study, 
we relax this assumption and consider all components of the magnetic field to show 
the role of nonzero vertical magnetic flux.}. With the help of the above 
assumptions, the three components of the current density, $ \bm{J} $, can be 
read as

\begin{equation} \label{J_r}
	J_{r} = \frac{1}{r \sin \theta} \frac{\partial}{\partial \theta} \left( B_{\phi} \sin \theta \right),
\end{equation}

\begin{equation} \label{J_theta}
	J_{\theta} = - \frac{1}{r} \frac{\partial}{\partial r} \left( r B_{\phi} \right),
\end{equation}

\begin{equation}
	J_{\phi} = 0.
\end{equation}

By considering toroidal component of the magnetic field, the anisotropic tensor 
can be reduced to

\begin{eqnarray}
\bm{\Pi}  & = &
 \left[
\begin{array}{ccc}
\Pi_{rr} & 0 & 0 \\
0 & \Pi_{\theta \theta} & 0 \\
0 & 0 & \Pi_{\phi \phi}
\end{array}  \right], 
\end{eqnarray}

with

\begin{multline}
\Pi_{\mathrm{rr}} = \Pi_{\theta \theta} = -\frac{1}{2} \Pi_{\phi \phi} 
= \rho \nu_{2} \Bigg[ \frac{v_{r}}{r} + \frac{v_{\theta}}{r} \cot\theta \\ 
- \frac{1}{3r^{2}} \frac{\partial}{\partial r} (r^{2} v_{r}) 
- \frac{1}{3r \sin \theta} \frac{\partial}{\partial \theta} (\sin \theta v_{\theta})  \Bigg].
\end{multline}

In the MHD case, thermal conductivity is mostly anisotropic being suppressed 
in the direction transverse to the magnetic field. The form of the thermal conduction 
can be written as

\begin{equation} \label{Q_conduction}
	\bm{Q}_\mathrm{c} = \kappa \nabla T,
\end{equation}
where $ T $ is the gas temperature and $ \kappa$ is the thermal conduction coefficient.
Although the saturated form of the thermal conductivity is appropriate for 
hot accretion flows we adopt a standard form of thermal conduction in which the heat flux depends 
linearly on the local temperature gradient. Consequently, the radial dependency of the 
thermal conductivity coefficient $ \kappa $ is considered to be a power-law function of radius to 
preserve radial self-similarity, i.e., $ \kappa(r) = \kappa_{_0} r^{1/2 - n} $, where 
$ n $ is the density index (see \citealt{Tanaka and Menou 2006} and 
\citealt{Khajenabi and Shadmehri 2013} for more details). 
By substituting all the above assumptions and definitions into Equations 
(\ref{eq:continuity})-(\ref{eq:delB}), we obtain partial differential equations 
(PDE) presented in Appendix \ref{pde_appendix}. 

\subsection{Self-similar Solutions} \label{subsec:self_similar_solutions}

Based upon the numerical HD and MHD simulations of hot accretion flow,
time-averaged inflow and outflow mass accretion rates decrease with radius due to the existence of
wind (\citealt{Stone et al. 1999}; \citealt{Yuan et al. 2012a, Yuan et al. 2012b}). Consequently,
their results show that density profile becomes flatter than the case with constant mass accretion rate,
in the range of $ 10 r_{s} < r < 100 r_{s} $ ( far away from the strong gravity of central black hole and 
also outer boundary conditions), where $ r_{s} $ is the 
Schwarzschild radius. In this study, following the numerical
simulations, we consider radial power-law form for the physical variables including density 
to remove the radial dependency of the variables. To do so, we adopt the following self-similar solutions in 
the radial direction by adopting a fiducial radius $ r_{\!_{0}} $. The solutions can be written as

\begin{equation}\label{rho_selfsimilar}
  \rho \left(r, \theta \right) = \rho_{_{0}} \left( \frac{r}{r_{_{0}}} \right)^{-n} \rho(\theta),
\end{equation}

\begin{equation}\label{vr_selfsimilar}
  v_{r} (r, \theta) = v_{_{0}} \left( \frac{r}{r_{_{0}}} \right)^{-1/2} v_{r}(\theta), 
\end{equation}

\begin{equation}\label{vtheta_selfsimilar}
  v_{\theta} (r, \theta) = v_{_{0}} \left( \frac{r}{r_{_{0}}} \right)^{-1/2} v_{\theta}(\theta),
\end{equation}

\begin{equation}\label{vphi_selfsimilar}
  v_{\phi} (r, \theta) = v_{_{0}} \left( \frac{r}{r_{_{0}}} \right)^{-1/2} v_{\phi}(\theta),
\end{equation}

\begin{equation}\label{c_s_selfsimilar}
  c_\mathrm{s} (r, \theta) = v_{_{0}} \left(  \frac{r}{r_{_{0}}} \right)^{-1/2} c_\mathrm{s} (\theta),
\end{equation}

\begin{equation} \label{Bp_selfsimilar}
	B_{\phi}(r,\theta) = B_{_0} \left( \frac{r}{r_{_{0}}} \right)^{-(n/2)-(1/2)} b_{\phi}(\theta),
\end{equation}
where $ r_{_{0}} $, $ \rho_{_{0}} $, $ v_{_{0}} \left[= \sqrt{GM/r_{_{0}}} \right]$, and 
$ B_{_{0}} \left[ = v_{_0} \sqrt{4 \pi \rho_{_0}} \,\right] $ are the units of length, density, 
velocity, and magnetic field, respectively. The ordinary differential equations (ODEs) 
can be derived by substituting the above self-similar solutions into the PDEs 
(\ref{pde_continuity})–(\ref{pde_induction}). The coupled system of ODEs is presented in 
Appendix \ref{ode_appendix}. The ODEs (\ref{ode_continuity})-(\ref{ode_induction}) 
consist of six physical variables: $ v_{r} (\theta) $, $ v_{\theta} (\theta) $, $ v_{\phi} (\theta) $, 
$ \rho(\theta) $, $ c_\mathrm{s}(\theta) $, $ b_{\phi}(\theta) $ as well as their derivatives. 

\subsection{Boundary Conditions}

All physical variables are assumed to be even symmetric, continuous, 
and differentiable at equatorial plane. Hence, $ v_{r}(\theta) = v_{r}(\pi - \theta) $,
$ v_{\theta} (\theta) = - v_{\theta}(\pi - \theta) $, $ v_{\phi}(\theta) = v_{\phi}(\pi - \theta) $,
$ c_\mathrm{s}(\theta) = c_\mathrm{s}(\pi - \theta) $, and $ \rho(\theta)  = \rho(\pi - \theta) $.
Since we include the latitudinal component of the velocity, $ v_{\theta} $, due to even 
symmetric assumption, its value will be zero at this boundary. 
Thus, the following boundary conditions will be imposed at $ \theta = \pi/2 $: 

\begin{equation}
	\frac{\mathrm{d} \rho}{\mathrm{d} \theta} = \frac{\mathrm{d} c_{s}}{ \mathrm{d} \theta} = 
	\frac{\mathrm{d} v_{r}}{\mathrm{d} \theta} = \frac{\mathrm{d} v_{\phi}}{ \mathrm{d} \theta} = 
	\frac{\mathrm{d} b_{\phi}}{\mathrm{d} \theta}  = v_{\theta} = 0.
\end{equation}

We define the plasma beta which is the ratio of the gas to the magnetic 
pressure. To study the models with weak and strong magnetic fields, 
we then set its value at the equatorial plane as

\begin{equation} \label{beta_at_midplane} 
	\beta_{_0} = \frac{p_\mathrm{gas}}{p_\mathrm{mag}} = \frac{2 \rho c_\mathrm{s}^{2}}{b_{\phi}^{2}}.
\end{equation}  

Here, the dimensionless 
magnetic pressure is defined as $ p_\mathrm{mag} = b_{\phi}^2 / 2 $. Since the 
maximum density is located at the equatorial plane, we adopt $ \rho(\pi/2) = 1$
for all set of input parameters. By substituting the above boundary conditions 
into Equations (\ref{ode_continuity})-(\ref{ode_induction}), we can get the
following equations at the equatorial plane

\begin{equation} \label{dvt_at_midplane}
	\frac{\mathrm{d} v_{\theta}}{\mathrm{d} \theta} = \left( n - \frac{3}{2} \right) v_{r},
\end{equation}

\begin{multline}  \label{mom1_at_midplane}
	- \frac{1}{2} v_{r}^{2} - v_{\phi}^{2} = -1 + \left( n + 1 + \frac{n - 1}{\beta} \right) c_\mathrm{s}^{2} \\
	- \alpha_{2} \left( n - 2 \right) \left(1 + \frac{1}{\beta} \right)  \left[ \frac{1}{2} v_{r} - \frac{1}{3} \frac{\mathrm{d} v_{\theta}}{\mathrm{d} \theta} \right] c_\mathrm{s}^{2},
\end{multline}

\begin{equation} \label{mom3_at_midplane}
	v_{r} = 3 \alpha_{1} \left( n - 2 \right) \left(1 + \frac{1}{\beta} \right) c_\mathrm{s}^{2},
\end{equation}

\begin{multline} 
	\left( n  - \frac{1}{\gamma - 1} \right) v_{r}  = \frac{9}{4} \alpha_{1} \left(1 + \frac{1}{\beta} \right) v_{\phi}^{2} \\
	+ \frac{1}{2} \eta_{_0} \left( n - 1 \right)^{2} \left(1 + \frac{1}{\beta} \right) c_\mathrm{s}^{2} + 3 \alpha_{2} \left(1 + \frac{1}{\beta} \right) \\
	\times \left[ \frac{1}{2} v_{r} -\frac{1}{3} \frac{\mathrm{d}v_{\theta}}{\mathrm{d} \theta} \right]^2 + \kappa_{_0} \left( n - \frac{1}{2} + \frac{1}{c_{s}^{2} }\frac{\mathrm{d}^{2} c_{s}^{2}}{\mathrm{d} \theta^{2}} \right).
\end{multline}

\begin{figure*}[ht!]
\includegraphics[width=0.95\textwidth]{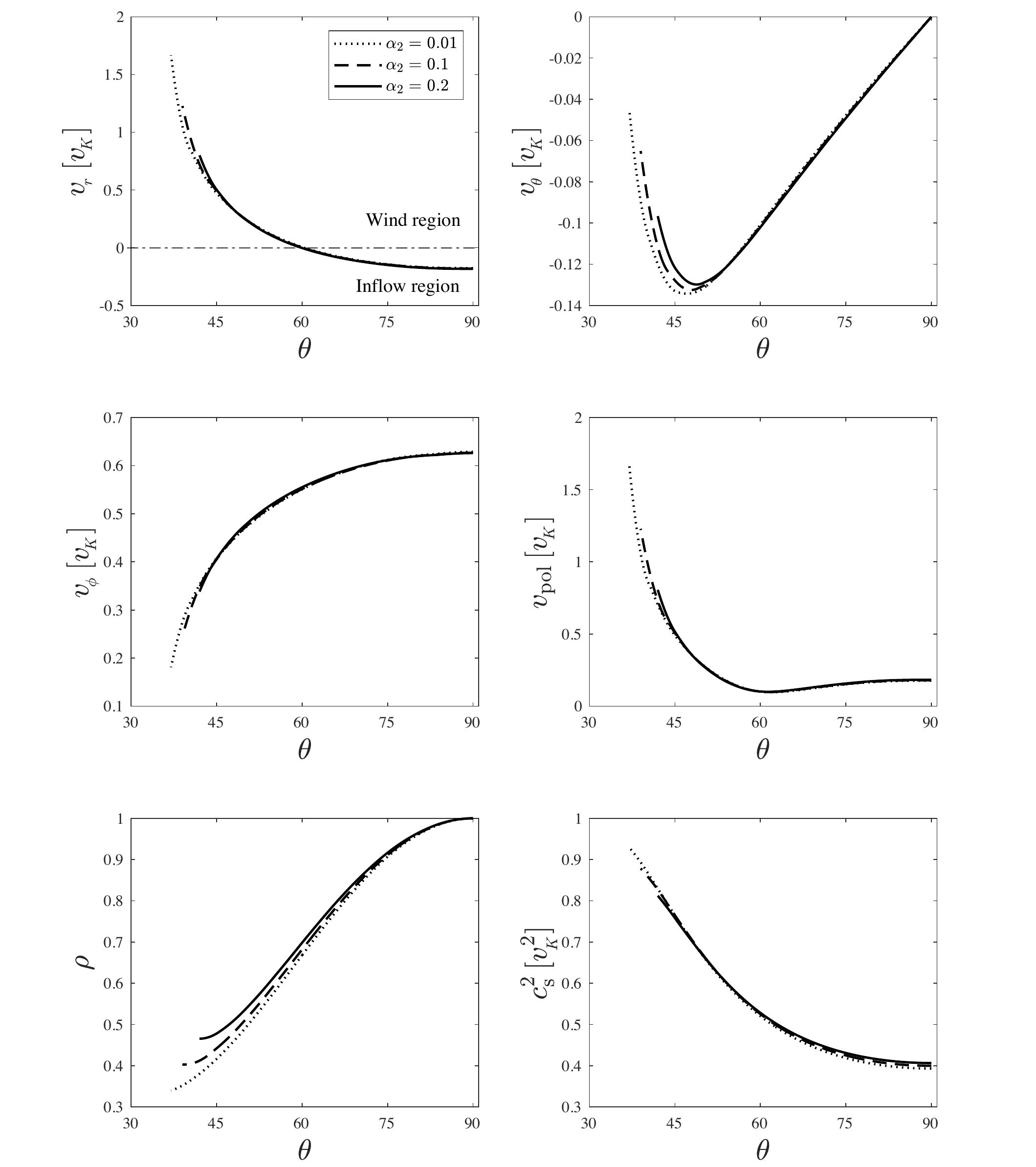}
\centering
\caption{Latitudinal profiles of the physical variables. Top left: radial velocity, $ v_{r} $, 
Top right: latitudinal velocity, $ v_{\theta} $, Middle left: angular velocity, $ v_{\phi} $,
Middle right: poloidal velocity, $ v_\mathrm{pol} = \sqrt{ v_{r}^{2} + v_{\theta}^2 } $, 
Bottom left: density, and Bottom right: sound speed squared, $ c_\mathrm{s}^{2} $. 
The velocities are in the unit of Keplerian velocity, $ v_{\!_K} $, and the density is in the 
unit of the density at the equatorial plane. The dotted, dashed, and solid curves 
correspond to parameters $ \alpha_{2} = [0.01, 0.1, 0.2] $, respectively. 
Here, $ \alpha_{1} = 0.1 $, $ n = 0.5 $, $ \gamma = 5/3 $, $ \eta_{_0} = 0.1 $, and 
$ \beta_{_0} = 5000 $.   
\label{variables1}}
\end{figure*}

\begin{figure*}[ht!]
\includegraphics[width=\textwidth]{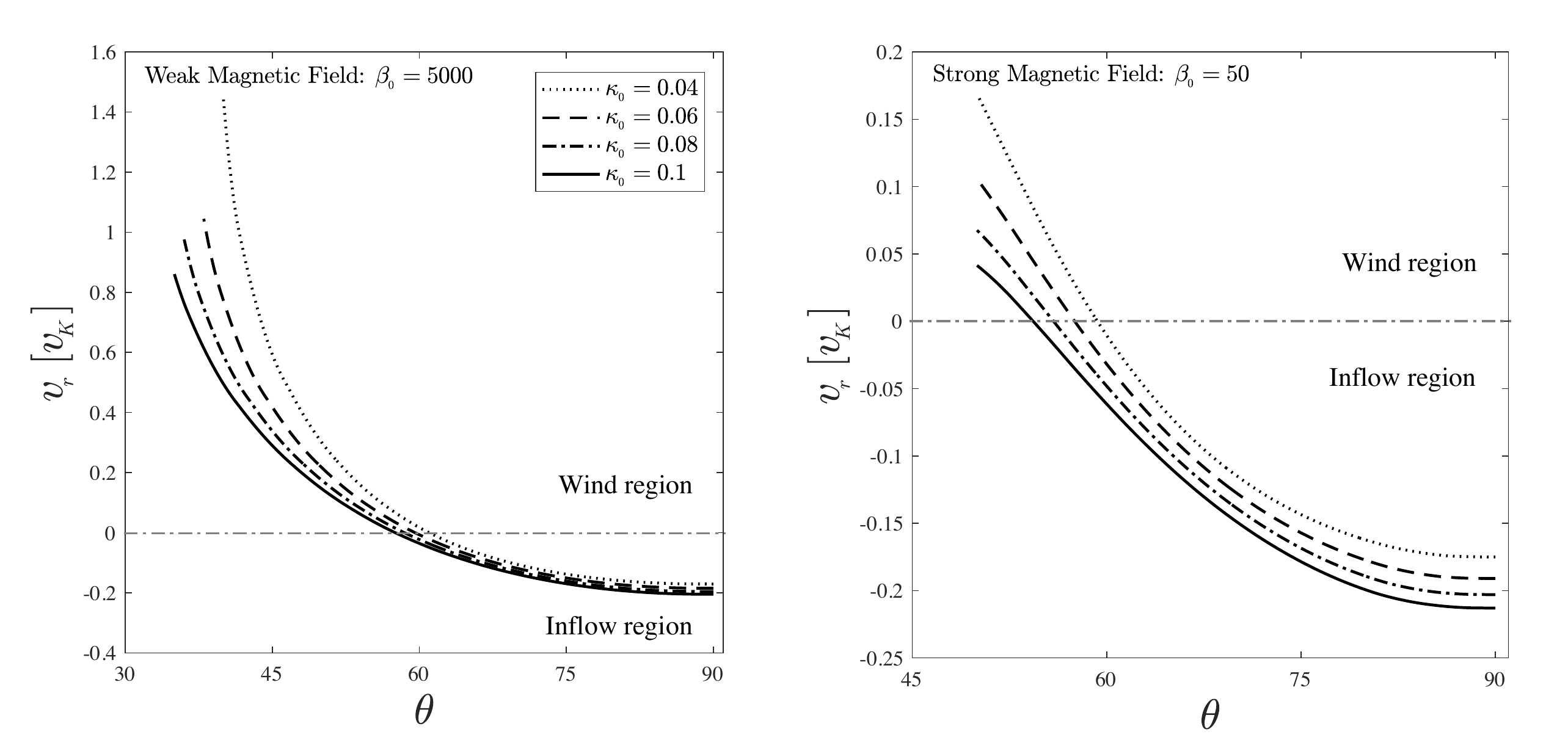}
\centering
\caption{Latitudinal profiles of the radial velocity for four different values of $ \kappa_{_0} = 
[0.04, 0.06, 0.08, 0.1] $, and for two cases of weak (left column) and strong (right column) 
magnetic fields. Here, $ \alpha_{1} = 0.1 $, $ \alpha_{2} = 0.05 $, $ n = 0.5 $, 
$ \gamma = 5/3 $, and $ \eta_{_0} = 0.1 $. 
\label{vr_beta0}}
\end{figure*} 

\begin{figure}[ht!]
\includegraphics[width=0.47\textwidth]{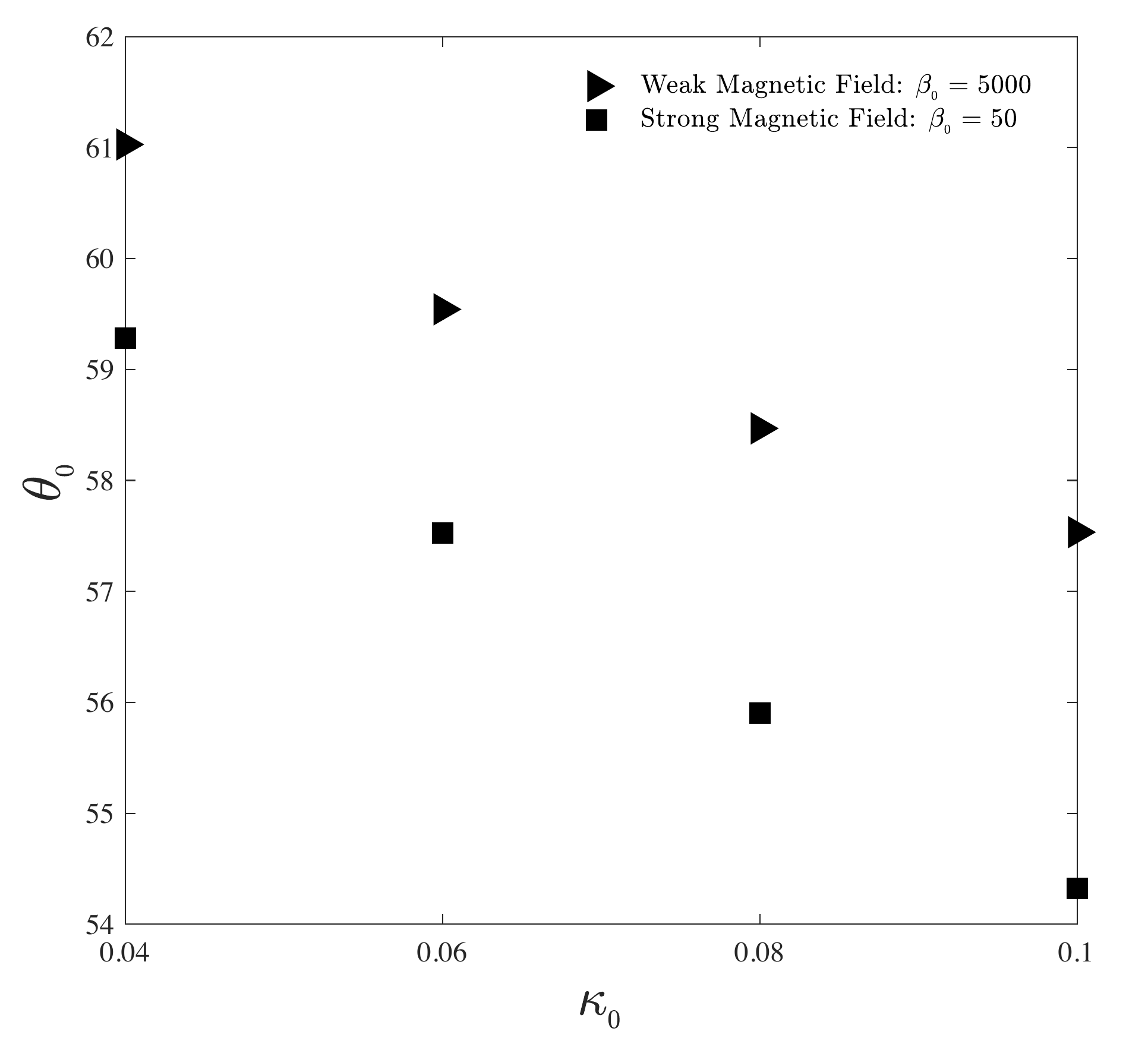}
\centering
\caption{Opening angle $ \theta_{_0} $ versus the thermal condition coefficient $ \kappa_{_0} $ 
with $ \alpha_{1} = 0.1 $, $ \alpha_{2} = 0.05 $, $ n = 0.5 $, $ \gamma = 5/3 $, and $ \eta_{_0} = 0.1 $. 
The two models with weak, $ \beta_{_0}  = 5000 $, and strong magnetic fields, $ \beta_{_0}  = 50 $, 
are marked by triangles and squares, respectively.   
\label{theta0}}
\end{figure}

It is noting that without thermal conduction, i.e., $ \kappa_{0} = 0 $, the above set of equations 
simply gives us the values of $ v_{r} $, $ v_{\phi} $, and $ c_\mathrm{s} $ at the 
equatorial plane. In the presence of the thermal conduction, we have another unknown
parameter, i.e., ($ \mathrm{d}^{2} c_\mathrm{s}^{2} / \mathrm{d} \theta^{2} $), at the 
equatorial plane. To have a self-consistent solutions, we follow \citealt{Khajenabi and Shadmehri 
2013} approach and guess the radial velocity at the equatorial plane. We can then 
simply obtain the value of other physical variables such as $ c_\mathrm{s} $, 
$ v_{\phi} $ and $ b_{\phi} $ from Equations (\ref{dvt_at_midplane})-(\ref{mom3_at_midplane})
and (\ref{beta_at_midplane}). We also impose the following physical 
boundary conditions at the angle where the radial velocity becomes null, 
i.e., $ v_{r} (\theta_{_0}) = 0 $,

\begin{equation} \label{wind_constraint}
	c_\mathrm{s}^{2}(\theta_{_0}) = \frac{\pi}{2} - \theta_\mathrm{0}.
\end{equation}

The above dimensionless constraint at $ \theta_{_0} $ represents this fact that 
the scale-height of accretion disk, $ H = c_\mathrm{s} / \Omega_{_K} $
should be equal to the disk height, i.e., $ c_\mathrm{s} / \Omega_{_K} \cong 
( \pi/2 - \theta_{_0}) r $. More precisely, the height of the disk is proportional 
to the wind temperature. By integrating the Equations (\ref{ode_continuity})-
(\ref{ode_induction}), we find the angle at which the radial velocity of the flow
becomes null. At this angle we check whether the wind condition 
(\ref{wind_constraint}) at $ \theta_{_0} $ is satisfied. If not, we choose another
value for the radial velocity at the equatorial plane until the Equation
(\ref{wind_constraint}) is satisfied. We then integrate the system of ODEs from the equatorial plane 
to a certain critical angle denoting as $ \theta_\mathrm{_B} $ where we encounter the numerical errors. 
More precisely, the mass density or the gas pressure will get very close to zero at this inclination.
We consider this angle as the upper boundary of the flow structure.
We think that with a simple radial self-similar solution we cannot describe the flow structure 
near the rotation axis. The reason is as follows: if the solution does not end at an upper boundary, 
$ \theta_\mathrm{_B} $, the net mass accretion rate $ \dot{M}_{\rm net} $, at a specific radius $ r $ 
is calculated as,

\begin{multline}
	\dot{M}_\mathrm{net} = \sqrt{GM} \left[ 4\pi \int^{_0}_{\pi/2} \rho(\theta) v_{r}(\theta)\sin\theta \mathrm{d} \theta \right] r^{3/2-n}\\
	= 4\pi \sqrt{GM} \Bigg[ \int^{\theta_{\!_0}}_{\pi/2} \rho(\theta) v_{r}(\theta)\sin\theta \mathrm{d} \theta \\
	+\int^{_0}_{\theta_{\!_0}} \rho(\theta) v_{r}(\theta)\sin\theta \mathrm{d} \theta \Bigg] r^{3/2-n} 
	= \dot{M}_{\rm in} + \dot{M}_{\rm out}.
\end{multline}

The mass conservation implies that net mass accretion rate 
should be a constant for a steady accretion flow. The constant mass 
accretion rate can only happen in the following two cases: (1) when 
$ n = 3/2 $, which forces $ \dot{M}_{\rm net} $ not to change with radius. 
This case is discussed in \citealt{Narayan and Yi 1995}, resulting a solution 
in which the flow is radial (with rotation), i.e., $ v_{\theta} = 0 $. 
Therefore, no outflow has been obtained in their solution.
(2) When $ n \neq 3/2 $, the integration term in the above equation 
must be zero, resulting the case $ \dot{M}_{\rm net} = 0 $. This requires 
that the outflow exactly equals the inflow at a certain radius, which 
in fact is not physical (no accretion process happens). Since we should 
have a non-zero net accretion rate, the inflow rate must be larger than 
the outflow rate at a certain radius, the solution must be truncated at 
a certain angle near the rotational axis, namely $ \theta_{_B} $. 
Consequently, the solution with outflow ($ n < 3/2 $) cannot satisfy self-similar approximations 
near the rotation axis.

By adopting this method, the outflow region extends to 
$ \theta_\mathrm{_B} < \theta < \theta_{_0} $. We think our solution in the region of 
$ \theta_\mathrm{_B} < \theta < \pi/2 $ is physical due to this fact that the solution satisfies 
the equations and boundary conditions at both ends as well as $ \theta = \theta_{_0} $.  
 
\section{Numerical Results} \label{sec:results}

\begin{figure*}[ht!]
\includegraphics[width=0.9\textwidth]{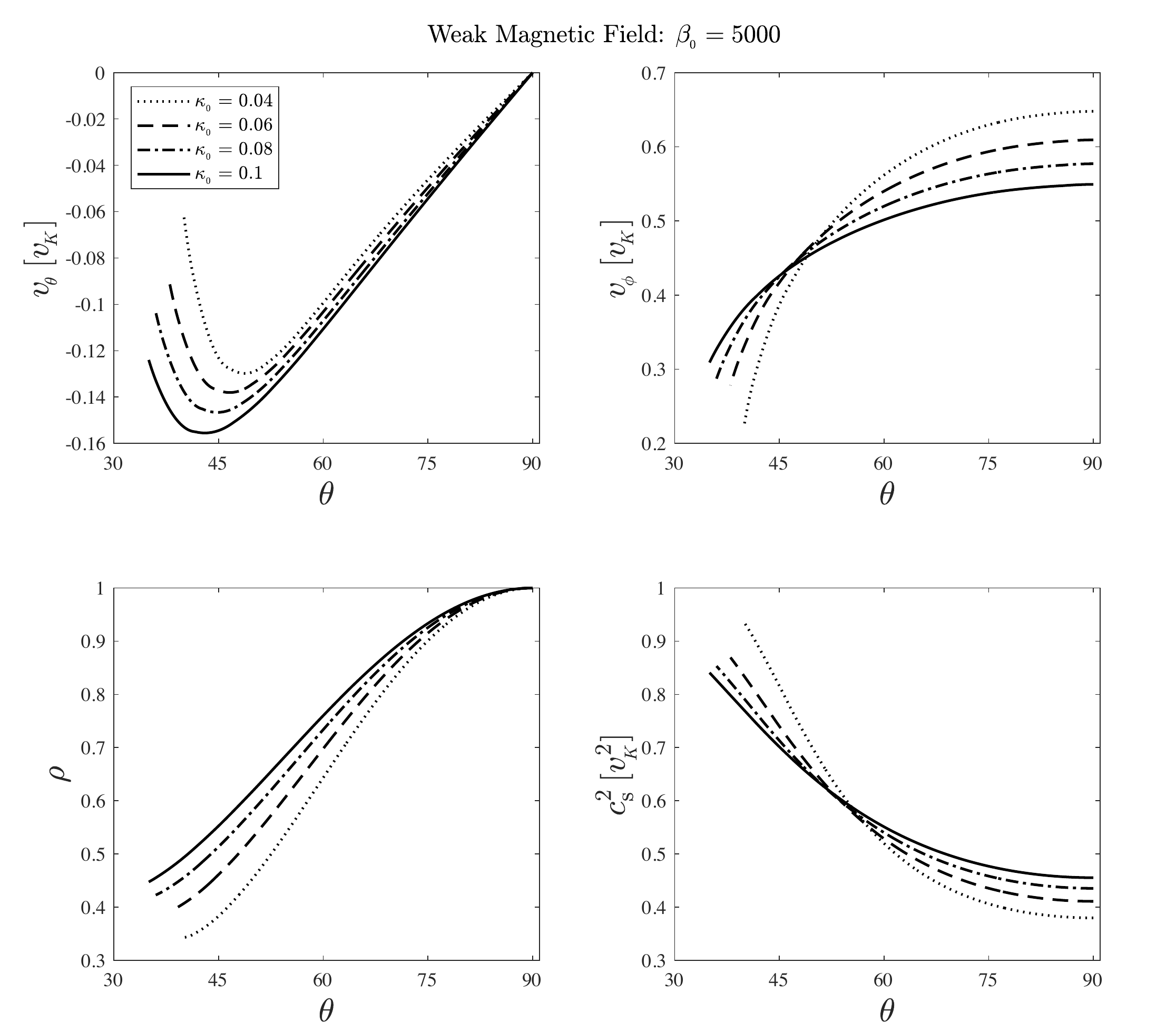}
\centering
\caption{Latitudinal profiles of the physical variables for weak magnetic field, $ \beta_{_0} = 5000 $. 
Top left: latitudinal velocity, : Top right: rotational velocity, $ v_{\phi} $, Bottom left: density, and 
Bottom right: sound speed squared, $ c_\mathrm{s}^{2} $. 
The dotted, dashed, dot-dashed and solid curves correspond to parameters $ \kappa_{_0} = [0.04, 0.06, 
0.08] $ and $ 0.1 $, respectively. Here, $ \alpha_{1} = 0.1 $, $ \alpha_{2} = 0.05 $, $ n = 0.5 $, 
$ \gamma = 5/3 $, and $ \eta_{_0} = 0.1 $.   
\label{weak_k0}}
\end{figure*}

\begin{figure*}[ht!]
\includegraphics[width=0.9\textwidth]{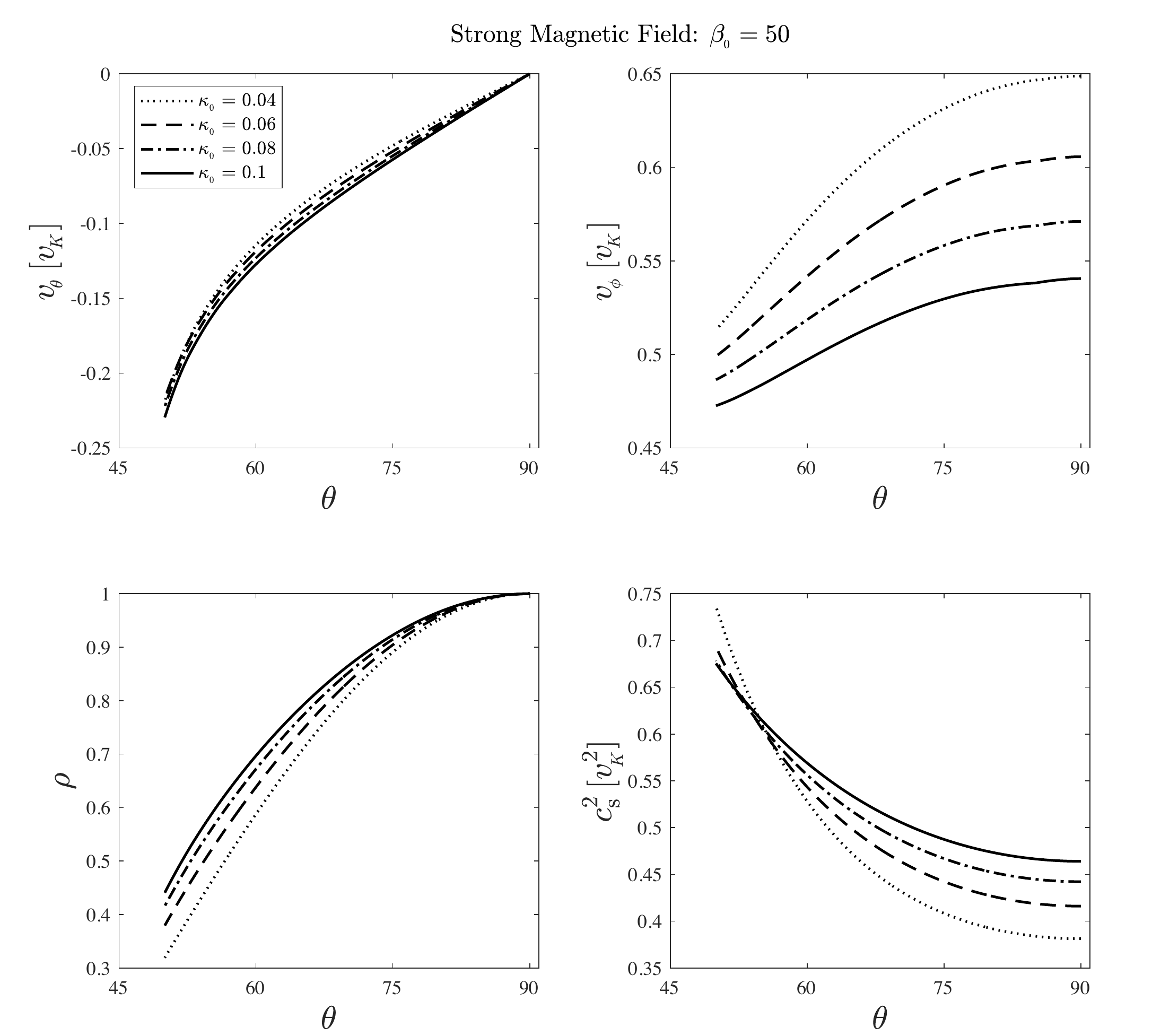}
\centering
\caption{Latitudinal profiles of the physical variables for strong magnetic field, $ \beta_{_0} = 50 $. 
Top left: latitudinal velocity, : Top right: rotational velocity, $ v_{\phi} $, Bottom left: density, and 
Bottom right: sound speed squared, $ c_\mathrm{s}^{2} $. 
The dotted, dashed, dot-dashed and solid curves correspond to parameters $ \kappa_{_0} = [0.04, 0.06, 
0.08] $ and $ 0.1 $, respectively. Here, $ \alpha_{1} = 0.1 $, $ \alpha_{2} = 0.05 $, $ n = 0.5 $, 
$ \gamma = 5/3 $, and $ \eta_{_0} = 0.1 $.   
\label{strong_k0}}
\end{figure*}

\begin{figure*}[ht!]
\includegraphics[width=\textwidth]{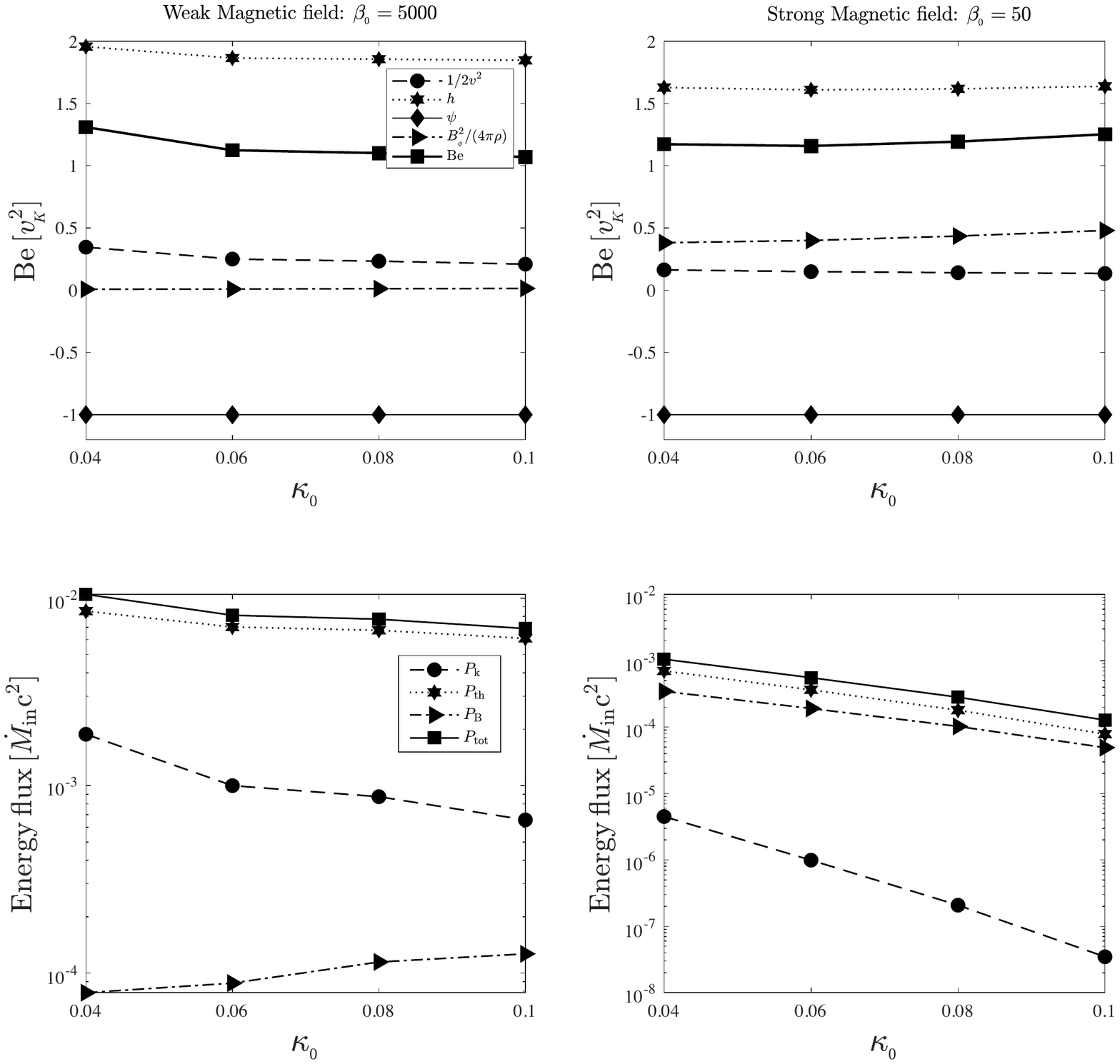}
\centering
\caption{Bernoulli parameter (top row panels) and energy flux (bottom row panels) for 
four different values of $ \kappa_{_0} = [0.04, 0.06, 0.08, 0.1] $, and for two cases of 
weak (left column) and strong (right column) magnetic fields. 
\label{Bernoulli}}
\end{figure*}

To avoid a shock being caused supersonic inflow such as jet near 
the central region, which is a source of deviation from the radial self-similar 
assumptions, we neglect the region within $ 10 r_{s} $. 
We consider radial power-law form for the physical 
variables to reduce the radial dependency of variables and solve the system 
of equations only in theta direction. Therefore, we believe our radial self-similar 
solutions are unique and stable. By this methodology, the PDEs 
are reduced to ODEs (Equations (\ref{ode_continuity})-(\ref{ode_induction})) 
which we solved numerically.
The equations were integrated from the equatorial plane, 
$ \theta = \pi/2 $, toward a critical angle, $ \theta = \theta_\mathrm{_B} $.
The solutions are in the following parameter space: $ \alpha_{1} = 0.1 $, 
$ \eta_{_0} = 0.1 $, $ \gamma = 5/3 $, and $ n = 0.5 $. To explore how much 
the anisotropic pressure can affect the physical variables of the accretion flow, 
we plotted Figure \ref{variables1} for three different values of $ \alpha_{2} = 
[0.01, 0.1, 0.2] $ in vertical direction and are shown by dotted, dashed, and 
solid lines, respectively. For this figure, $  \beta_{_0} = 5000 $ which represents 
weak magnetic field case \footnote{We also examin anisotropic pressure effects
for strong magnetic field case and the results were almost identical to Figure 
\ref{variables1}}. From left to right and top to bottom, it is shown the 
radial velocity, $ v_{r} $, latitudinal velocity, $ v_{\theta} $, the rotational velocity, 
$ v_{\phi} $, the poloidal velocity, $ v_{\mathrm{pol}} $, the density, $ \rho $, 
and the sound speed squared, $ c_{s}^2 $, respectively. All velocities are 
plotted in the unit of Keplerian value and the density is in the unit of maximum 
density at the equatorial plane at that radius. It is clear that the anisotropic 
pressure does not considerably affect the physical quantities; however, it can 
decrease the amount of the wind. For instance, in the top left panel, the wind 
region extends from the dot-dashed line that is $ v_{r} = 0 $ line by around 
$ \theta \sim 35^\circ $ for $ \alpha_{2} = 0.01 $ while for $ \alpha_{2} = 0.2 $ 
this region shrinks and goes until $ \theta \sim 43^\circ $. As it was easily 
expected, the radial velocity of the wind is much higher than that of the inflow. 
In the top right panel, there is an increase in the absolute value of the latitudinal 
velocity which the maximum amount occurs between $ \theta \sim 45^{\circ} - 50^{\circ} $. 
Then, we can see that the absolute value of this quantity decreases
toward the polar region. Therefore, the poloidal velocity of the flow shows 
an increasing trend from the equatorial plane toward the polar region. 
In the middle left panel, it is shown that $ v_{\phi} $ increases with $ \theta $.
The bottom left panel illustrates the density drops from the equatorial plane
toward the polar region. In addition, the anisotropic pressure can increase
the density especially in the wind region. This implies that as the amount of 
the wind is enhanced with decreasing $ \alpha_{2} $, density decreases due to 
the sweeping more gas away by the wind. As you can see in the bottom right 
panel of figure \ref{variables1}, the sound speed has an increasing behavior 
from the disk toward the polar region. 

We have not only examined the effects of the anisotropic pressure but 
we also investigated the influence of the thermal conductivity and the 
magnetic field strength on the flow properties in Figures \ref{vr_beta0}, 
\ref{weak_k0}, and \ref{strong_k0}. In the left and right panels of Figure 
\ref{vr_beta0}, the radial velocity is plotted for weak and strong magnetic 
field with $ \beta = [5000, 50] $, respectively. We also consider four different 
values for the conductivity coefficient as $ \kappa_{_0} = [0.04, 0.06, 0.08, 0.1] $ 
illustrated by dotted, dashed, dot-dashed, and solid lines, respectively. 
This figure clearly shows that in the weak magnetic field, thermal conduction 
does not affect the inflow radial velocity. However, in the case of the strong 
magnetic field thermal conduction can increase the inflow radial velocity. 
In addition, thermal conduction can decrease the radial wind velocity 
disregarding the magnetic field strength. 

Figure \ref{theta0} shows the dependency of $ \theta_{_0} $
on $ \kappa_{_0} $ for two models with weak and strong magnetic fields,
$ \beta = [5000, 50] $ marked by triangles and squares, respectively. 
The inverse behavior of these two parameters can be clearly seen 
from this figure, $ \theta_{_0} $ decreases when $ \kappa_{_0} $ increases. 
As $ \theta_{_0} $ is the declination angle between inflow and wind areas, 
increasing $ \kappa_{_0} $ leads to a gradual shrinkage of the wind region. 
For the weak magnetic field the wind region is expanded respect to the 
strong magnetic field. For instance, for $ \kappa = 0.06 $ the wind region 
in weak magnetic field starts from $ \theta \sim 61^{\circ}$ while in strong 
magnetic field it starts from $ \theta \sim 59^{\circ}$.

The impact of the thermal conduction on other physical variables of the accretion 
flow in two weak and strong magnetic field models, $ \beta = [5000, 50] $, is 
examined in Figures \ref{weak_k0} and \ref{strong_k0}, respectively. Latitudinal velocity in 
the unit of Keplerian value is plotted in the top left panel of Figure \ref{weak_k0}. 
Although the thermal conduction does not show any effects on $ v_{\theta} $ 
around the equatorial plane, it causes the absolute value of wind latitudinal 
velocity increases. In the case of the strong magnetic field in Figure \ref{strong_k0}, 
the absolute value of the latitudinal velocity drastically increases toward 
the polar region. In the top right panel of Figure \ref{weak_k0} it is seen that
when magnetic field is weak, thermal conductivity shows a different effect on the 
rotational velocity of the inflow and wind. In the inflow, the thermal conductivity 
reduces the rotational velocity however it can increase the wind velocity.
We cannot see this inverse behavior in the inflow and wind region when the 
magnetic field magnifies. In the case of strong magnetic field, thermal conduction 
causes a decrease in the rotational velocity of both inflow and wind (top right panel 
of Figure \ref{strong_k0}). In both cases of the magnetic field the density increases 
with the thermal conduction and this implies that the thermal conduction can make 
the disk thicker. In addition, in the weak magnetic field the density drops smoother 
rather than the case with strong magnetic field (bottom left panels of Figures 
\ref{weak_k0} and \ref{strong_k0}). The sound speed squared in the unit of Keplerian 
velocity is plotted in the bottom right panel 
of both Figures \ref{weak_k0} and \ref{strong_k0}. A comparison between 
these two panels shows that thermal conduction can reduce the sound speed 
and thus the temperature of the flow. Further, it can be seen that for the higher 
values of the thermal conduction the sound speed increases smoothly from the 
equatorial plane toward the polar region. More, when the magnetic field strength 
increases the sound speed decreases in the wind region.

The Bernoulli parameter with only considering toroidal magnetic field is defined as
\begin{equation}
	\mathrm{Be} = \frac{1}{2} v^2 + h + \psi + \frac{B_{\phi}^2}{4\pi \rho}
\end{equation}
where $ h (= \gamma p / [\rho (\gamma - 1)]) $ is enthalpy. The top row panels 
of Figure \ref{Bernoulli} show the Bernoulli parameter for four values of the 
conductivity coefficients, $ \kappa = [0.04, 0.06, 0.08, 0.1] $, and also for weak 
and strong magnetic fields, $ \beta = [5000, 50] $, respectively. As you can see, 
enthalpy is higher in the weak magnetic field with respect to the strong magnetic 
field. However, we cannot see significant changes in the Bernoulli parameter
when the strength of the magnetic field is amplified.

We also calculated the wind power which includes kinetic, thermal, 
and Poynting energy fluxes as

\begin{equation}
	P_{\mathrm{k}}(r) = 2 \pi r^2 \int^{90^\circ}_{0^\circ} \rho \, \mathrm{max}(v_r^{3} , 0) \sin \theta \mathrm{d}\theta,
\end{equation}
\begin{equation}
	P_{\mathrm{th}}(r) = 4 \pi r^2 \int^{90^\circ}_{0^\circ} \rho e \, \mathrm{max}(v_r , 0) \sin \theta \mathrm{d}\theta,
\end{equation}
\begin{equation}
	P_{\mathrm{B}}(r) = 4 \pi r^2 \int^{90^\circ}_{0^\circ} \mathrm{S_{r}} \, \mathrm{max}(v_r/|v_r| , 0) \sin \theta \mathrm{d}\theta.
\end{equation}

where $ \mathrm{S_{r}} $ is the radial component of Poynting flux and is defined as
\begin{equation}
\mathrm{S_{r}} = v_{r} \frac{B_{\phi}^2}{4\pi}.
\end{equation}

The results are plotted in the bottom row of Figure \ref{Bernoulli} with respect to the 
conductivity coefficient where the left and right panels are for weak and strong magnetic fields, 
respectively. All components of the energy flux decrease with the conductivity coefficient in both
cases of the magnetic field except for the Poynting energy flux in the weak magnetic field which 
increases with this parameter. Clearly, the thermal energy flux dominates two other fluxes 
in both weak and strong magnetic field cases. Moreover, when the magnetic field is weak
the kinetic energy flux is larger than the Poynting energy flux. As the magnetic field magnifies, 
the Poynting energy flux prevails over the kinetic energy.

\section{Summary and Discussion}\label{sec:summary_discussion}

In this paper, we solve the equations of the hot accretion flow by considering
magnetic field, anisotropic pressure and thermal conduction. We investigate
the properties of the inflow and wind of the collissionless plasmas in hot accretion
flows with very low mass accretion rates. As increase or decrease in the 
magnetic field strength makes a pressure anisotropy, we compare 
the physical quantities of the accretion flow in two cases with weak  
and strong magnetic field. We decompose the magnetic field into a large-scale 
component and a turbulent component. We only consider toroidal component of 
magnetic field. The effects of turbulence are described by the viscosity. 
Both components can transfer the angular momentum outward and produce heat.
We modeled the anisotropic pressure with an anisotropic viscosity.
By considering the self-similar solutions, we solved the ODEs numerically.
We examined the anisotropic pressure on the properties of the inflow and wind.
Our results showed that the increase of the anisotropic pressure
tends to give rise to shrink the wind region. We also investigate 
the effects of the thermal conduction and the magnetic field strength 
on the properties of the inflow and wind. We found that the impact 
of the thermal conductivity on the flow to some extent depends on 
the magnetic field strength. As in the strong magnetic field, 
thermal conduction tends to decrease the rotational velocity of the wind.
Thermal conductivity also tends to shrink the wind region. Our results 
showed that neither thermal conduction nor the magnetic field strength 
can impressively impact on the Bernoulli parameter. We explored 
the energy flux of the wind in details to know how thermal conduction 
and the magnetic field strength can affect the thermal, kinetic, and 
Poynting energy fluxes. We found that, except the Poynting energy flux, 
the thermal conduction can decrease all components of the energy flux
in both weak and strong magnetic fields. What is more, the strength 
of the magnetic field can help the Poynting energy flux overcomes 
the kinetic energy flux.
In terms of black hole mass, the main application of our hot accretion results is in the dim black hole 
sources, including the supermassive black hole in our Galactic center, low-luminosity 
active galactic nuclei (AGNs), and quiescent and hard states of 
black hole X-ray binaries. Additional physics, namely, radiative cooling 
and self-gravity can play a significant role to construct realistic physical models. 
These additional physics may not necessarily allow a self-similar solution. 
But in some cases, they can help to determine some of the parameters of the self-similar model.

\begin{acknowledgments}
A.M. is supported by the China Postdoctoral Science Foundation (grant No. 2020M673371).
F.Z.Z. is supported by the National Natural Science Foundation of China (grant No. 12003021) 
and also the China Postdoctoral Science Foundation (grant No. 2019M663664). 
L.M. is supported by the National Natural Science Foundation of China (grant No. 12171385).  
A.M. also acknowledges the support of Dr. X. D. Zhang at the Network Information 
Center of Xi’an Jiaotong University. The computation has made use of the High Performance 
Computing (HPC) platform of Xi’an Jiaotong University.
\end{acknowledgments}


%\vspace{5mm}
%\facilities{HST(STIS), Swift(XRT and UVOT), AAVSO, CTIO:1.3m,


%\software{astropy \citep{2013A&A...558A..33A,2018AJ....156..123A},  
%          Cloudy \citep{2013RMxAA..49..137F}, 
%          Source Extractor \citep{1996A&AS..117..393B}
%          }



\appendix

\section{Partial Diffenatal Equations in Spherical Coordinates}  \label{pde_appendix}

To simplify the set of equations (\ref{eq:continuity})-(\ref{eq:delB}), we consider the flow 
is in steady-state and axisymmetric ($ \partial/\partial t = \partial/\partial \phi = 0 $). 
Spherical polar coordinate system ($ r,\theta,\phi $) is adopted. We assume all 
components of the velocity field, $ \bm{v} = v_{r} \hat{r} + v_{\theta} \hat{\theta} + 
v_{\phi} \hat{\phi} $. We further assume only the toroidal component of  magnetic 
field is presented. By substituting all assumptions and definitions introduced in 
subsection \ref{subsec:basic_equations} into equations (\ref{eq:continuity})-(\ref{eq:delB}),
we obtain following partial differential equations (PDE). Hence, the continuity equation 
is reduced to the following form,

\begin{equation}\label{pde_continuity}
  \frac{1}{r^{2}} \frac{\partial}{\partial r} \left( r^{2} \rho v_{r} \right) + \frac{1}{r \sin \theta} 
  \frac{\partial}{\partial \theta} \left( \rho v_{\theta} \sin\theta \right) = 0.
\end{equation}

The three components of the momentum equation are as,

\begin{equation}\label{pde_mom1}
\rho \left[ v_{r} \frac{\partial v_{r}}{\partial r}  + \frac{v_{\theta}}{r} \left( \frac{\partial v_{r}}{\partial \theta} - v_{\theta} \right) - \frac{v_{\phi}^{2}}{r} \right] = - \frac{GM \rho}{r^{2}} - \frac{\partial p_\mathrm{gas}}{\partial r}  + \frac{J_{\theta} B_{\phi}}{4 \pi} + \frac{1}{r^{2}}\frac{\partial}{\partial r} (r^{2}\Pi_{rr}) - \frac{1}{r} (\Pi_{\theta \theta} + \Pi_{\phi \phi}),
\end{equation}

\begin{equation}\label{mom2}
 \rho \left[  v_{r} \frac{\partial v_{\theta}}{\partial r} + \frac{v_{\theta}}{r} \left( \frac{\partial v_{\theta}}{\partial \theta} + v_{r} \right) - \frac{v_{\phi}^{2}}{r} \cot \theta \right] = - \frac{1}{r} \frac{\partial p_\mathrm{gas}}{\partial \theta} - \frac{J_{r} B_{\phi}}{4 \pi} + \frac{1}{r \sin \theta} \frac{\partial}{\partial \theta} (\sin \theta \Pi_{\theta \theta})
- \frac{\Pi_{\phi \phi}}{r} \cot \theta,
\end{equation}

\begin{equation}\label{mom3}
   \rho \left[ v_{r} \frac{\partial v_{\phi}}{\partial r} + \frac{v_{\theta}}{r} \frac{\partial v_{\phi}}{\partial \theta} + \frac{v_{\phi}}{r} \left( v_{r} +  v_{\theta} \cot\theta \right)  \right] =   \frac{1}{r^{3}} \frac{\partial}{\partial r} \left( r^{3} \sigma_{r \phi} \right).
\end{equation}

The energy equation is driven as,

\begin{multline}\label{pde_energy}
	\rho \left( v_{r} \frac{\partial e}{\partial r}   + \frac{v_{\theta}}{r}  \frac{\partial e}{\partial \theta} \right) - \frac{p_\mathrm{gas}}{\rho} \left( v_{r} \frac{\partial \rho}{\partial r}   + \frac{v_{\theta}}{r}  \frac{\partial \rho}{\partial \theta} \right)  = \sigma_{r \phi} r \frac{\partial}{\partial r} (\frac{v_{\phi}}{r} ) + \frac{\eta}{4 \pi} \left( J_{r}^{2} + J_{\theta}^{2}  \right) + \frac{1}{r^2} \frac{\partial}{\partial r} \left( r^2 \kappa \frac{\partial T}{\partial r} \right) \\
	+ \frac{1}{r \sin\theta}\frac{\partial}{\partial \theta} \left( \frac{\kappa \sin\theta}{r} \frac{\partial T}{\partial \theta} \right) - \Pi_{rr} \frac{\partial v_{r}}{\partial r} - \frac{\Pi_{\theta \theta}}{r} \left( \frac{\partial v_{\theta}}{\partial \theta} + v_{r} \right) -\frac{\Pi_{\phi \phi}}{r} (v_{r} + v_{\theta} \cot \theta).
\end{multline}

Finally, the induction equation will be written as follows,

\begin{equation}\label{pde_induction}
\frac{\partial}{\partial r} (r v_{r} B_{\phi}) +\frac{\partial}{\partial \theta} (v_{\theta} B_{\phi})
-\frac{\partial}{\partial \theta} (\eta J_{r}) + \frac{\partial}{\partial\theta} (r \eta J_{\theta}) = 0.
\end{equation}

\section{Ordinary Differential Equations} \label{ode_appendix}

The ordinary differential equations (ODEs) can be derived just by substituting 
self-similar solutions presented in section \ref{subsec:self_similar_solutions} 
into the partial differential equations (\ref{pde_continuity})–(\ref{pde_induction}) as,

\begin{equation} \label{ode_continuity}
	 \rho \left[ \left(\frac{3}{2} - n \right) v_{r} + \frac{\mathrm{d} v_{\theta}}{\mathrm{d} \theta} 
	+ v_{\theta} \cot \theta \right] + v_{\theta} \frac{\mathrm{d} \rho}{\mathrm{d}\theta} = 0,
\end{equation}

\begin{equation} \label{ode_mom1}
	\rho \left[ - \frac{1}{2} v_{r}^{2} + v_{\theta} \frac{\mathrm{d} v_{r}}{\mathrm{d} \theta} - v_{\theta}^{2} - v_{\phi}^{2} \right]
	= - \rho + \left( n + 1 \right) p_\mathrm{gas} +  j_{\theta} b_{\phi} - \alpha_{2} p_\mathrm{tot} (n - 2) \left[ \frac{1}{2} v_{r} + \frac{2}{3} v_{\theta} \cot \theta - \frac{1}{3} \frac{\mathrm{d} v_{\theta}}{\mathrm{d} \theta}  \right],
\end{equation}

\begin{multline} \label{ode_mom2}
	\rho \left[ \frac{1}{2} v_{r} v_{\theta} + v_{\theta} \frac{\mathrm{d} v_{\theta}}{\mathrm{d} \theta} - v_{\phi}^{2} \cot \theta \right] = 
	- \frac{\mathrm{d} p_\mathrm{gas}}{\mathrm{d} \theta} - j_{r} b_{\phi} + \alpha_{2} \left( \frac{\mathrm{d} p_\mathrm{tot}}{\mathrm{d} \theta} + 3 p_\mathrm{tot} \cot \theta \right) \left[ \frac{1}{2} v_{r} + \frac{2}{3} v_{\theta} \cot \theta - \frac{1}{3} \frac{\mathrm{d} v_{\theta}}{\mathrm{d} \theta} \right] \\	
	+ \alpha_{2} p_\mathrm{tot} \left[ \frac{1}{2} \frac{\mathrm{d} v_{r}}{\mathrm{d} \theta} 
	+ \frac{2}{3} \left( \frac{\mathrm{d} v_{\theta}}{\mathrm{d} \theta} \cot \theta - v_{\theta} \csc^{2} \theta \right) 
	- \frac{1}{3} \frac{\mathrm{d}^{2} v_{\theta}}{\mathrm{d} \theta^{2}} \right],
\end{multline}

\begin{equation}\label{ode_mom3}
\rho \left[ \frac{1}{2} v_r v_{\phi} + v_{\theta} \frac{\mathrm{d} v_{\phi}}{\mathrm{d}\theta} + v_{\theta} v_{\phi} \cot \theta \right] =
 \frac{3}{2} \alpha_{1} \left(n - 2 \right) p_\mathrm{tot} v_{\phi},
\end{equation}

\begin{multline}\label{ode_energy}
\left( n  - \frac{1}{ \gamma - 1} \right) v_{r}  p_\mathrm{gas} + \frac{v_{\theta}}{\gamma - 1}  \left( \frac{\mathrm{d} p_\mathrm{gas}}{\mathrm{d}\theta} - \gamma \frac{p_\mathrm{gas}}{\rho} \frac{\mathrm{d} \rho}{\mathrm{d} \theta} \right) = \frac{9}{4} \alpha_{1} p_\mathrm{tot} v_{\phi}^{2} + \eta\left( j_{r}^{2} + j_{\theta}^{2} \right) \\
+ 3 \alpha_{2}\, p_\mathrm{tot} \left( \frac{1}{2} v_{r} + \frac{2}{3} v_{\theta} \cot \theta -\frac{1}{3} \frac{\mathrm{d}v_{\theta}}{\mathrm{d} \theta} \right)^2 + \kappa_{_0} \left[ \left( n - \frac{1}{2} \right) c_{s}^{2} + \frac{1}{\sin\theta} \frac{\mathrm{d}}{\mathrm{d} \theta} \left(\sin \theta \frac{\mathrm{d} c_{s}^{2}}{\mathrm{d} \theta} \right)  \right],
\end{multline}

\begin{equation}\label{ode_induction}
\frac{n}{2}v_{r}b_{\phi} - v_{\theta} \frac{\mathrm{d}b_{\phi}}{\mathrm{d}\theta}
-b_{\phi}\frac{\mathrm{d}v_{\theta}}{\mathrm{d}\theta} + \eta \frac{\mathrm{d}j_{r}}{\mathrm{d}\theta}
+ j_{r} \frac{\mathrm{d}\eta}{\mathrm{d}\theta} + \frac{n}{2} \eta j_{\theta} = 0,
\end{equation}
where

\begin{equation} \label{ode_pgas}
	p_\mathrm{gas} = \rho c_{s}^{2} 
\end{equation}

\begin{equation} \label{ode_ptot}
	p_\mathrm{tot} = p_\mathrm{gas} + p_\mathrm{mag} 
\end{equation}

\begin{equation} \label{ode_pmag}
	p_\mathrm{mag} = \frac{b_{\phi}^{2}}{2} 
\end{equation}

\begin{equation} \label{ode_jr}
	j_{r} = \frac{\mathrm{d} b_{\phi}}{\mathrm{d} \theta} + b_{\phi} \cot \theta,
\end{equation}

\begin{equation} \label{ode_jt}
	j_{\theta} = \frac{1}{2} \left( n - 1 \right) b_{\phi},
\end{equation}

\begin{equation} \label{ode_eta}
	\eta = \eta_{_0} \frac{p_\mathrm{tot}}{\rho}.
\end{equation}


\bibliography{aniso}{}
\bibliographystyle{aasjournal}



\end{document}